\documentclass{article}

\usepackage[margin=25mm]{geometry}
\usepackage{amsmath}
\usepackage{amsfonts}
\usepackage{amssymb}
\usepackage{graphicx}
\usepackage{verbatim}
\usepackage{soul}
\usepackage{lineno}
\usepackage{cite}
\usepackage{subfig}
\usepackage{siunitx}

\newcommand\blfootnote[1]{%
  \begingroup
  \renewcommand\thefootnote{}\footnote{#1}%
  \addtocounter{footnote}{-1}%
  \endgroup
}
\DeclareSIUnit\gauss{G}
\usepackage[utf8]{inputenc} 
\title{\textbf{Laser frequency stabilization by modulation transfer spectroscopy and balanced detection of molecular iodine for laser cooling of ${}^{174}$Yb }}

\author{Álvaro M.G. de Melo$^{1}$, Hector Letellier$^{1}$, Apoorva Apoorva$^{1}$, Antoine Glicenstein$^{1}$, Robin Kaiser$^{1,*}$   \\
        \small \textit{${}^{1}$Université Côte d’Azur}, CNRS, INPHYNI, 17 Rue Julien Lauprête, 06200 Nice, France\\
}

\begin{document}

\maketitle
\blfootnote{${}^{*}$robin.kaiser@univ-cotedazur.fr}


\begin{abstract}
We report laser frequency stabilization by the combination of modulation transfer spectroscopy and balanced detection of a relatively weak hyperfine transition of the R(158)25-0 line of molecular iodine (${}^{127}$I$_{2}$), which is used as a new frequency reference for laser trapping and cooling of ${}^{174}$Yb on the $^{1}S_{0} - {}^{3}P_{1}$ transition. The atomic cloud is characterized by time-of-flight measurements, and an on-resonance optical depth of up to 47 is obtained. We show laser noise reduction and characterize the short-term laser frequency instability by the Allan deviation of the laser fractional frequency. The minimum measured value is 3.9 \ifmmode\times\else\texttimes\fi $10^{-13}$ at \SI{0.17}{\second} of averaging time. 
\end{abstract}

\section{Introduction}

Atoms with two valence electrons, i.e. alkaline-earth like metals, have attracted the attention of the laser cooling community since the end of the last century \cite{Shimizu,Ertmer,Potterveld,Willmann,Yabuzaki} and their applications have ranged from precision measurements \cite{Tino} to quantum computing \cite{Zoller}. Among them, strontium and ytterbium have been the species often chosen due to their internal structure allowing the use of a dipole-allowed transition at \SI{461}{\nano\meter} and \SI{399}{\nano\meter}, respectively, to slow down atoms from the atomic source in combination with a narrow closed transition ($^{1}S_{0} - {}^{3}P_{1}$) of linewidths $2 \pi \cdot 7$ \SI{}{\kilo \hertz} and $2 \pi \cdot 182$ \SI{}{\kilo \hertz}. These narrow transitions can be used to achieve high optical depths at low temperatures.

Compared to broad line magneto-optical traps (MOT) such as Rubidium with a $2 \pi \cdot 6$ \SI{}{\mega \hertz} linewidth and easy-to-use spectroscopic absorption cells, one challenge in realizing an ytterbium MOT operating on the $^{1}S_{0} - {}^{3}P_{1}$ intercombination line is to stabilize the laser frequency, due to its linewidth ($\Gamma_{g} = 2\pi \cdot 182$ \SI{}{\kilo \hertz}) and the difficulty of realizing an appropriate Yb spectroscopic cell \cite{Gupta} to perform usual saturated absorption spectroscopy on the $^{1}S_{0} - {}^{3}P_{1}$ transition. 
As $\Gamma_{g}$ is below commercial wavemeter's precision\cite{Weidemuller}, different techniques were used by several groups to circumvent such difficulties. Frequency modulated (FM) spectroscopy of the atomic beam has been used in \cite{Cornish}, resulting in a short-term frequency stability of less than \SI{1}{\mega \hertz}. Other groups have stabilized the laser to a cavity line that is locked to another stable reference \cite{Fortson, Lu, Takahashi}. Such methods have shown to be efficient, however dependent on the atomic flux or cavity stability.

Similarly to what is done for alkalis using a gas cell to perform spectroscopy, molecular iodine ($^{127}$I$_{2}$), has also been used to stabilize the laser used to load an Yb MOT on the $^{1}S_{0} - {}^{3}P_{1}$ transition\cite{Dareau}. The frequency reference was a convenient $^{127}$I$_{2}$ hyperfine transition in the P(49)24-1 rotational line. The required frequency shift to reach the \SI{1.3}{\giga \hertz} shifted $^{174}$Yb $^{1}S_{0} - {}^{3}P_{1}$ transition has been realized using three acousto-optic modulators (AOMs), two of them in double-pass configuration. In this work we show that a simple configuration can be implemented by choosing a different $^{127}$I$_{2}$ transition and combining the robustness of modulation transfer spectroscopy (MTS) with respect to temperature and laser power fluctuations to the common-noise rejection achieved by balanced detection. The details on the experimental scheme, spectroscopy of molecular iodine and laser frequency stabilization feedback loop are given in section \ref{Experiment}. In section \ref{Results}, we present the implementation of an Yb MOT with 2 \ifmmode\times\else\texttimes\fi $10^{8}$ atoms operating on the $^{1}S_{0} - {}^{3}P_{1}$ intercombination transition that was obtained with the laser stabilized by the proposed method. We also give details on the laser noise reduction and characterize the short-term instability of the laser frequency by means of the Allan deviation of the laser's fractional frequency.

\section{Laser frequency stabilization scheme and iodine spectroscopy}\label{Experiment}
Among the various laser locking (i.e. frequency stabilization) techniques available, we have chosen the so-called modulation transfer spectroscopy (MTS), where the pump beam is modulated in a saturated absorption spectroscopy configuration.
In this technique the modulation of the pump field is transferred to the probe by a four-wave mixing (FWM) process as explained in \cite{Ducloy}. MTS is generally chosen due to its robustness to temperature drifts and laser intensity fluctuations \cite{DucloyMTS}. Moreover, the increase of frequency stability by MTS in a balanced detection, i.e. when the saturated absorption profile is subtracted by a linear absorption signal as depicted in Fig.\ref{MTSScheme}, was reported in \cite{Mio}. Therefore, we employ MTS in a balanced configuration to stabilize the laser frequency to a relatively weak $^{127}$I$_{2}$ line that is convenient for laser cooling of $^{174}$Yb atoms on the $^{1}S_{0} - {}^{3}P_{1}$ line.

\subsection{Experimental scheme}
\begin{figure}[h!]
\centering
\includegraphics[width=12cm]{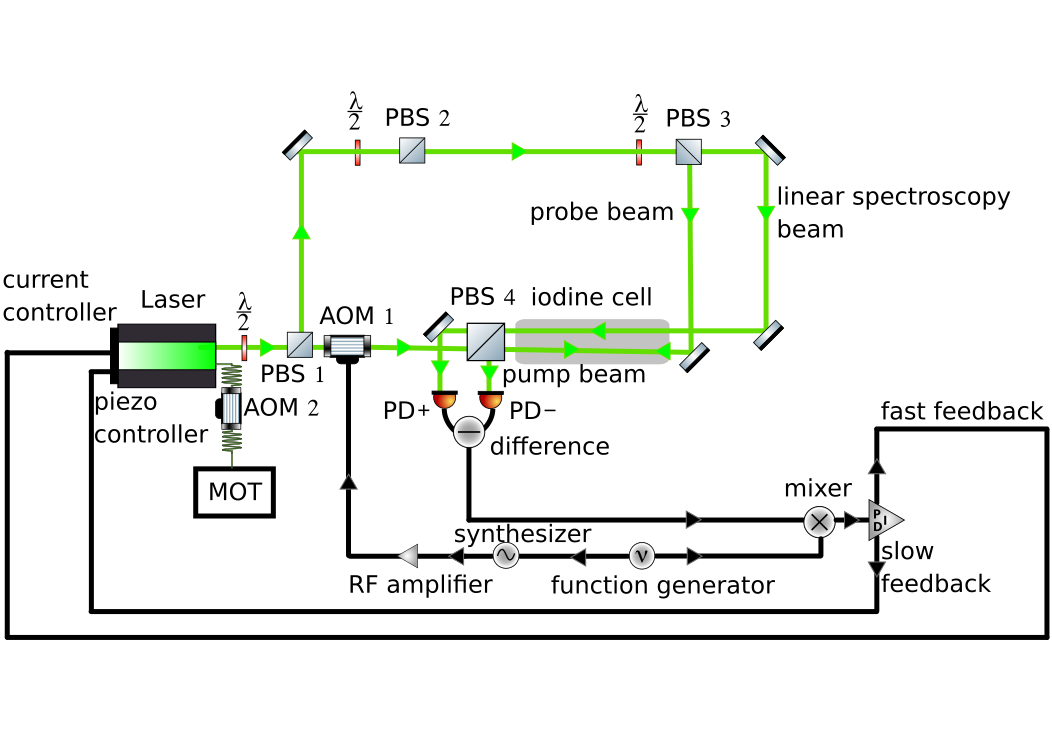}
\caption{Simplified experimental scheme for the frequency stabilisation of the laser used to load a MOT on the $^{174}$Yb $^{1}S_{0} - {}^{3}P_{1}$ line. The laser beam is split into two inside the laser system, both outputs (a fibered and a free space) are shown. The free space output is divided into two different beams by PBS 1. One beam passes through AOM 1 where it is diffracted and modulated to be used as the pump on the MTS, while the other optical beam is further split into two by PBS 3. The reflection of PBS 3 is the probe for the MTS and the transmission is the beam used to perform linear spectroscopy of $^{127}$I$_{2}$. The difference between the MTS probe and the linear absorption is sent to a RF mixer to be demodulated to generate the error signal that is sent as a feedback to the laser controllers. The fibered output is diffracted by AOM 2 before being split into six by a fiber splitter before being sent to the glass cell to perform the MOT.}
\label{MTSScheme}
\end{figure}

The laser system (Toptica TA-SHG pro) consists of a diode laser that generates around \SI{50}{\milli \watt} at \SI{1112}{\nano \meter} before being amplified by a tapered amplifier to up to \SI{1.8}{\watt} and finally a second-harmonic generation (SHG) cavity that delivers up to \SI{1.3}{\watt} at around \SI{556}{\nano \meter}. After the SHG cavity, the beam is divided into two by inserting a half-wave plate and a polarizing beam splitter (PBS) inside the laser box (not depicted). In Fig.\ref{MTSScheme}, it can be seen that one beam is sent to a free space output and the other one to a fiber output. The fiber output is connected to a fibered AOM (AOM 2, Brimrose TEM-150-9-60-556-2FP-HP), that is used to shift the laser frequency near resonance to the $^{174}$Yb $^{1}S_{0} - {}^{3}P_{1}$ transition as well as to provide time control of the amplitude of the MOT beams. After AOM 2, the beam is further split by a 2$\times$6 fiber splitter to generate the six MOT beams. The free space output depicted is used for laser frequency stabilization via MTS on $^{127}$I$_{2}$ using AOM 1 to perform the necessary frequency modulation for MTS. Therefore, the laser beams used for the MOT do not suffer from any residual frequency modulation, in contrast to the case when the laser diode current is modulated.

The transmission of PBS 1 is diffracted and modulated by AOM 1, and the -1 diffracted order is used as the pump beam for saturated absorption spectroscopy. The pump beam at the position of the cell has a power of \SI{300}{\milli \watt} and a full width at $1/e^{2}$ of \SI{0.75}{\milli \meter}. AOM 1 is driven by a direct digital synthesizer (2023A Aeroflex). The synthesizer's carrier frequency (\SI{80}{\mega \hertz}) is modulated by a \SI{200}{\kilo \hertz} sinusoidal signal generated by a function generator (B$\&$K Precision 4052) with a modulation index of 10. It is important to note that the -1 diffracted order of AOM 1 includes separated beams of frequency shifts defined by the driver's carrier frequency ($\Delta_{1} = $ \SI{-80}{\mega \hertz}) and the sidebands (\SI{-79.8}{\mega \hertz} and \SI{-80.2}{\mega \hertz}, for the first sidebands) generated by frequency modulation of this carrier signal \cite{Himsworth}. As the AOM is positioned close enough to the spectroscopic cell the carrier and the first sidebands are almost superposed along their propagation through the 10-cm spectroscopic cell. The transmission of PBS 2 is further split into two beams by another half-wave plate and PBS 3. The transmission is used to perform linear spectroscopy of $^{127}$I$_{2}$ and is measured by the input PD+ of a free space balanced photodetector (Thorlabs PDB210A/M). The reflection of PBS 3 is aligned counterpropagating to the pump beam and plays the role of the probe beam in saturated absorption spectroscopy, which is measured by the second input (PD-) of the photodetector. The spectroscopic cell is slightly tilted to avoid undesired pump beam reflections reaching PD- and is kept at room temperature without any cooling or heating system. Each probe beam has a power of \SI{1}{\milli \watt} and a full width at $1/e^{2}$ of \SI{0.75}{\milli \meter} at the spectroscopic cell. 

The second output of the function generator is sent to the local oscillator (LO) input of the RF mixer to perform the demodulation of the subtraction of the signal measured by PD+ and PD-. Finally, it is amplified and demodulated to generate an error signal that is sent to a PID controller (FALC 110), which has a high bandwidth ($\sim$ \SI{10}{\mega \hertz}) output and a slow integrator output with a bandwidth of \SI{10}{\kilo \hertz}. The slow integrator output is used to suppress long-term drifts and is fed back to the voltage applied to a piezo actuator holding the grating (FALC's slow integrator output). The fast feedback closes the loop controlling the laser diode's current (FALC's high bandwidth output).

\subsection{Iodine spectroscopy}
\begin{figure*}[h!]
    \begin{minipage}{.5\textwidth}
        \vspace*{\fill}
        \centering
        \subfloat[\label{LargeScan}]{\includegraphics[width=1.0\linewidth]{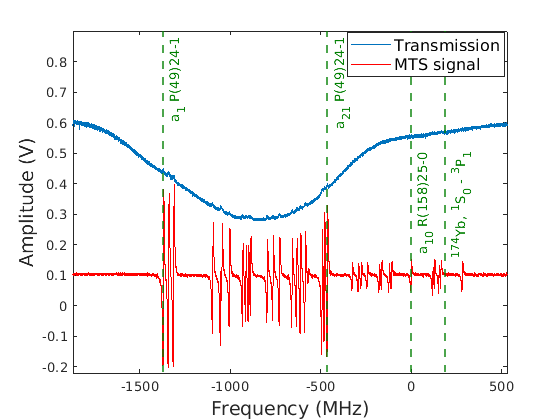}}
    \end{minipage}%
    \begin{minipage}{.5\textwidth}
        \vspace*{\fill}
        \centering
        \subfloat[\label{MidScan}]{\includegraphics[width=1.0\linewidth]{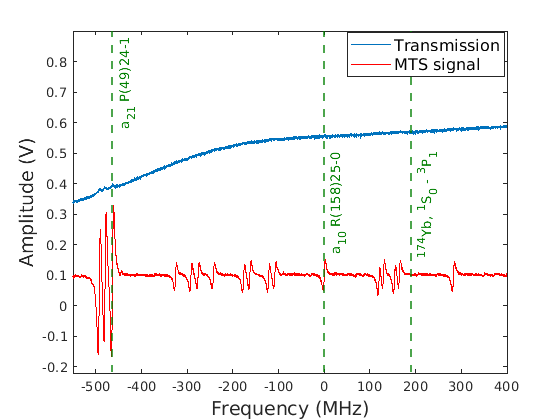}}
    \end{minipage}\\
    \centering
    \begin{minipage}{.5\textwidth}
        \vspace*{\fill}
        \centering
        \subfloat[\label{SmallScan}]{\includegraphics[width=1.0\linewidth]{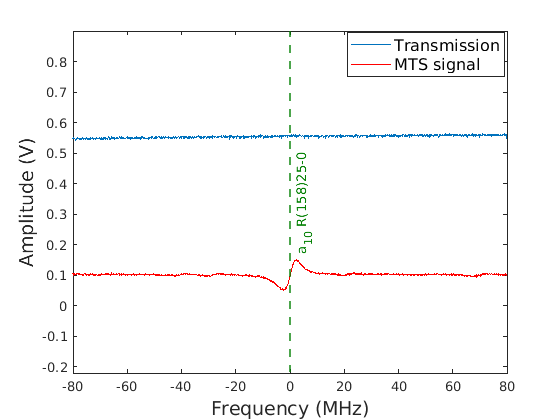}}
    \end{minipage}
    \caption{ \textbf{(a)} Saturated absorption spectroscopy of P(49)24-1 and R(158)25-0 $^{127}$I$_{2}$ lines (in blue) and error signals generated by MTS (in red) for a laser scan rate of \SI{147}{\mega \hertz \per \second}. The origin of the frequency axis is the center of the $a_{10}$ R(158)25-0 transition. The strong lines between \SI{-1500}{\mega \hertz} and \SI{-400}{\mega \hertz} are the 21 hyperfine transitions of the P(49)24-1 line. The weak lines shown from \SI{-400}{\mega \hertz} to \SI{300}{\mega \hertz} are 14 of the hyperfine transitions of the R(158)25-0 line. The green vertical dashed lines indicates the $a_{1}$ and $a_{21}$ lines of P(49)24-1, the $a_{10}$ line of R(158)25-0 and the $^{174}$Yb $^{1}S_{0} - {}^{3}P_{1}$ intercombination transition \textbf{(b)} Zoom of (a) from \SI{-550}{\mega \hertz} to \SI{1800}{\mega \hertz}. \textbf{(c)} Zoom of (a) from \SI{-80}{\mega \hertz} to \SI{80}{\mega \hertz}.    \label{SaturatedAbs}}
\end{figure*}

The saturated absorption signal measured by PD- (in blue) and the MTS error signal (in red) are shown in Fig.\ref{SaturatedAbs} for a laser frequency scan rate of \SI{147}{\mega \hertz \per \second}. The strong lines between \SI{-1500}{\mega \hertz} and \SI{-400}{\mega \hertz} in Fig.(\ref{LargeScan}) correspond to the 21 hyperfine components of the P(49)24-1 rotational line, one of them already used as a frequency reference for laser cooling of Ytterbium atoms as reported in \cite{Dareau}. Precision spectroscopy of the P(49)24-1 $^{127}$I$_{2}$ line was recently performed and reported in \cite{Hong}, where the absolute frequencies of the hyperfine transitions were measured using an optical frequency comb with an uncertainty of \SI{7}{\kilo \hertz} and with frequency shifts due to laser power (\SI{-2.6}{\kilo \hertz \per \milli \watt}) and iodine pressure (\SI{-1.8}{\kilo \hertz \per \pascal}). In Fig.\ref{SaturatedAbs}, we have used these results in order to calibrate our frequencies assuming a linear scanning of the frequency of our laser and chose the center of our frequency axis 
at the $a_{10}$ R(158)25-0 hyperfine line (set at \SI{0}{\mega \hertz}  on Fig.\ref{SaturatedAbs}). We choose to stabilize the frequency of our laser on the $a_{10}$ hyperfine transition of the R(158)25-0 line, defining thus a frequency reference for our laser used for laser cooling of $^{174}$Yb atoms as presented in the next section. The choice of this line is motivated by the fact that it is a mere \SI{191}{\mega \hertz} away from the  $^{174}$Yb ${}^{1}S_{0}-{}^{3}P_{1}$ frequency. The frequency-difference of the pump beam in the MTS by AOM 1 ($\Delta_1$) with respect to the probe beam shifts the frequency of the saturation peak by $\Delta_1 / 2 =$ \SI{-40}{\mega \hertz} \cite{DucloyShift}. We use the +1 diffraction order for AOM 2 with a frequency-shift of $\Delta_{2} =$ \SI{151}{\mega \hertz}. Combining the frequency shifts due to AOM 1 and AOM 2 we thus shift the laser frequency near the $^{174}$Yb resonance, with a total frequency-shift with respect to the $a_{10}$ hyperfine transition of the R(158)25-0 line of $\Delta = \Delta_{2} - \Delta_1 / 2 =$ \SI{191}{\mega \hertz}.

\begin{figure}[h!]
\centering
\includegraphics[width=0.5\linewidth]{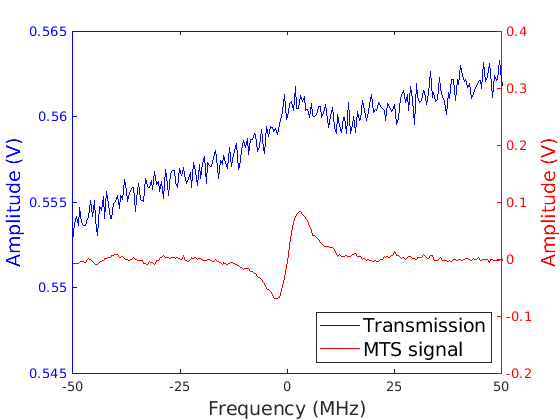}
\caption{The saturated absorption spectroscopy of the $a_{10}$ hyperfine component of the R(158)25-0 line, measured by PD-, is shown in blue for a laser frequency scan rate of \SI{6.65}{\giga \hertz \per \second}. The data shown in blue is an average of 100 samples and stresses the small relative amplitude of the line. The MTS error signal (in red) measured at the output of the FALC for the same laser frequency scan rate but without averaging.}
\label{SlowScan}
\end{figure}

Fig.\ref{MidScan} is a zoom of Fig.\ref{LargeScan} from \SI{-550}{\mega \hertz} to \SI{400}{\mega \hertz} that allows to compare the amplitude of the $a_{10}$ hyperfine component of the R(158)25-0 line with respect to the hyperfine components of the P(49)24-1 line. Fig.\ref{SmallScan}, a zoom of Fig.\ref{LargeScan} from \SI{-80}{\mega \hertz} to \SI{80}{\mega \hertz}, is centered at the $a_{10}$ hyperfine component of the R(158)25-0 line. The small amplitude of this line with respect to the P(49)24-1 hyperfine lines is probably the main reason why it was not yet exploited as a frequency reference for experiments using the ${}^{1}S_{0}-{}^{3}P_{1}$ transition of $^{174}$Yb atoms, despite being closer to $^{174}$Yb transition. In Fig.\ref{SlowScan}, we show a scan around the saturated absorption peak, highlighting the small amplitude of the absorption signal (in blue). 
The relative amplitude of the line, i.e. the amplitude of the peak with respect to the background divided by the background amplitude, is smaller than 1\%, meaning that it is below the laser's power stability over seconds. This is the main reason why MTS is more suitable to our experiment than frequency-modulation (FM) spectroscopy. In FM spectroscopy, the background of the error signal depends on fluctuations of the spectroscopic signal resulting in fluctuations of the error signal in time. In MTS the background level is very robust to signal's fluctuation \cite{Cho}, MTS is also very suitable to perform wideband locking as reported in \cite{Turner}. MTS has been widely used with acousto-optic modulators performing the pump beam frequency modulation by modulating the carrier frequency of the AOM driver \cite{Himsworth,Lin}. In red, we show the MTS error signal (without averaging) after being amplified by the FALC, which is the signal sent as a feedback to the laser's current and piezo controllers as well as the signal used to measure the laser frequency instability (considering the gain applied by the FALC) that will be shown in the next section. 

Absolute frequency measurements of the $^{176}$Yb ${}^{1}S_{0}-{}^{3}P_{1}$ line were reported in \cite{Natarajan,Mcferran}, with isotope shifts also reported allowing to connect to the frequency of the $^{174}$Yb ${}^{1}S_{0}-{}^{3}P_{1}$ line. Even though the goal of the present paper is not to perform an absolute frequency measurement of the $^{174}$Yb ${}^{1}S_{0}-{}^{3}P_{1}$ line, we note that the \SI{40}{\mega \hertz} discrepancy between the two reported values in \cite{Natarajan,Mcferran} can be addressed by our work with sufficient precision. Indeed, relying on the absolute frequency $\nu_{a21}$ of the $a_{21}$ hyperfine transition of P(49)24-1 $^{127}$I$_{2}$ line reported in \cite{Hong} ($\nu_{a21}=$ \SI{539 385 938 687}{\kilo \hertz}) and exploiting the measured absolute frequencies of the $a_{1}$ and $a_{21}$ hyperfine transitions of P(49)24-1 
$^{127}$I$_{2}$ lines to calibrate our scanning rate, we get for the frequency $\nu_{a10}$ of the $a_{10}$ R(158)25-0 hyperfine line an absolute  frequency of 
$\nu_{a10}=$ \SI{539 386 406 (6)}{\mega \hertz}, with an estimated precision of \SI{6}{\mega \hertz} in our frequency scan. 
Adding our AOM induced frequency shift of 191MHz (taking into account the laser detuning of
\SI{-1.1}{\mega \hertz})  we obtain for the frequency of the ${}^{1}S_{0}-{}^{3}P_{1}$ line of $^{174}$Yb a value of 
$\nu_{174}=$ \SI{539 386 597 (6)}{\mega \hertz}. This is consistent within the precision of our estimation with the value of  \SI{539 386 600}{\mega \hertz} reported in \cite{Mcferran} but in discrepancy with the value of  \SI{539 386 560}{\mega \hertz} reported in \cite{Natarajan}.

\section{Results}\label{Results}
\subsection{Magneto-optical trap}

In this section we present the realization of a MOT of $^{174}$Yb operating on the narrow ${}^{1}S_{0}-{}^{3}P_{1}$ transition, illustrating the convenient frequency stabilization of our laser to the $a_{10}$ hyperfine component of the R(158)25-0 line. Neutral ytterbium atoms are generally laser cooled by making use of two transitions, the one at \SI{399}{\nano \meter} is dipole-allowed and has a broad linewidth ($\Gamma_{b}= 2 \pi \cdot$ \SI{29}{\mega \hertz}) which makes it suitable for slowing down and capturing a large range of velocity classes and it is, therefore, 
the transition used for Zeeman slowers. The narrow transition can then be used in a second cooling step due to its narrower linewidth ($\Gamma_{g} = 2\pi \cdot$ \SI{182}{\kilo \hertz}) that allows to obtain colder atomic clouds with a Doppler temperature of \SI{4}{\micro \kelvin} for the bosonic isotopes (i.e. in the absence of sub-doppler cooling). The absence of any Zeeman degeneracy in ground state of the bosonic isotopes of Yb makes them suitable to perform experiments aiming at verify the existence of 3D localization of light as proposed in \cite{Skipetrov}.
In our experiment, as explained in \cite{Letellier}, we use an oven  with an array of micro-tubes to produce a collimated beam of Yb. We trap the atoms on a MOT operating on the broad ${}^{1}S_{0}-{}^{1}P_{1}$ transition, which allows, due to its larger natural linewidth ($\Gamma_{b}$), to capture atoms in a larger range of velocity classes from the atomic beam. To further increase the number of atoms captured by the MOT we make use of a slowing beam, i.e. a laser beam counterpropagating to the atomic beam, but without a Zeeman slower. With this system, we obtain up to 1 \ifmmode\times\else\texttimes\fi $10^{9}$ atoms in a MOT operating on the ${}^{1}S_{0}-{}^{1}P_{1}$ transition at \SI{399}{\nano \meter}  (blue MOT). This blue MOT is loaded using \SI{72}{\milli \watt} of total laser power split in three pairs of counterpropagating beams of waist \SI{22}{\milli \meter} and detuned by $-$2$\Gamma_{b}$, a slowing beam of power \SI{200}{\milli \watt} and detuned by $-$9.6 $\Gamma_{b}$ and an axial magnetic field gradient of \SI{20}{\gauss \per \centi \meter} for \SI{1}{\second}.

\begin{figure*}[h!]
    \begin{minipage}{.5\textwidth}
        \vspace*{\fill}
        \centering
        \subfloat[\label{MOTImages}]{\includegraphics[width=1.0\linewidth]{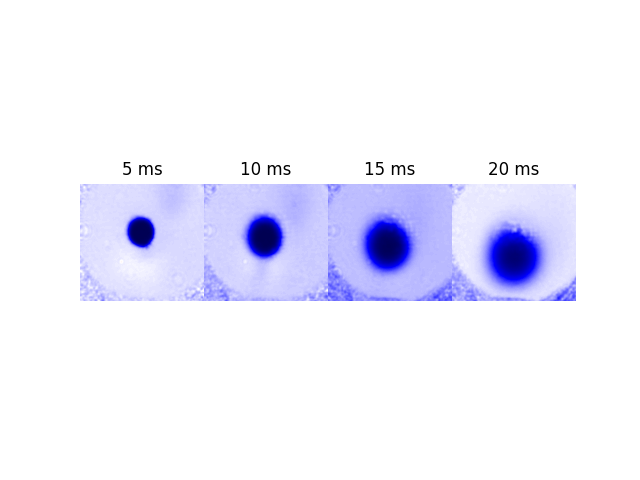}}
    \end{minipage}%
    \begin{minipage}{.5\textwidth}
        \vspace*{\fill}
        \centering
        \subfloat[\label{TOF}]{\includegraphics[width=1.0\linewidth]{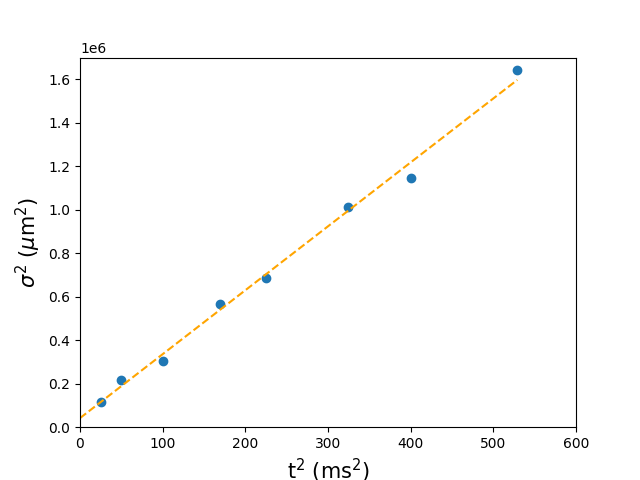}}
    \end{minipage}\\
    \caption{ (a) Absorption images of the atomic cloud for times of flight of 5, 10 ,15 and \SI{20}{\milli \second}. (b) Square of the rms width of the MOT along an horizontal axis as a function of the square of time of flight (blue dots) and a fit of the data (in orange). The number of atoms is 2 \ifmmode\times\else\texttimes\fi $10^{8}$, the MOT width at t=0 ms is \SI{262}{\micro \meter}, the on-resonance optical depth is $b_0=47$ and the temperature calculated from the curve fit is \SI{60}{\micro \kelvin} (with a  $R^{2}$ coefficient of the fit of 0.99). Error bars are not included as they are too small to be visible at this scale.  \label{MOTImages}}
\end{figure*}

To transfer the atoms from this blue MOT to our MOT on the narrow intercombination line, we reduce the axial magnetic field gradient to \SI{4}{\gauss \per \centi \meter} and turn off the blue MOT and slowing beams. The \SI{555.8}{\nano \meter} laser beam has a waist of \SI{9}{\milli \meter}, \SI{5}{\milli \watt} of total power split in the three pairs of counterpropagating beams, and is detuned by $-$6 $\Gamma_{g}$ from atomic resonance. We turn this lasers on for \SI{40}{\milli \second} to transfer the atoms to the MOT operating in the ${}^{1}S_{0}-{}^{1}P_{1}$ transition (green MOT). We characterize this green MOT by a time of flight technique and absorption imaging on the ${}^{1}S_{0}-{}^{1}P_{1}$ transition. We therefore turn off all laser beams and the magnetic field gradient during a time of flight (TOF) at which the atomic cloud experiences a free expansion and fall. After this time of flight, absorption images are obtained by a weak probe beam resonant with the ${}^{1}S_{0}-{}^{1}P_{1}$ transition. Images for TOF of 5, 10, 15, and \SI{20}{\milli \second} are shown in Fig.\ref{MOTImages}(a), illustrating the free fall of the cloud in the grativational field and its expansion due to its residual temperature. A quantitative analysis is done by plotting the square of the rms width of the MOT ($\sigma$) as a function of the square of the TOF (see Fig. \ref{MOTImages}(b) (blue circles)). The fit (dashed orange curve) of the data by the equation \cite{Shaffer,Gawlik}
\begin{equation}
    \sigma^{2}(t) = \frac{k_{B}T}{m}t^{2} + \sigma_{0}^{2},
\end{equation}
where $m$ is the atomic mass and $k_{B}$ is the Boltzman constant, allows us to extract the temperature of the cloud ($T\sim$ \SI{60}{\micro \kelvin}) and its intial size at $t=$ \SI{0} ($\sigma_{0} \sim$ \SI{262}{\micro \meter} along the horizontal axis). We also retrieve the number of atoms from these absorption images with $N \sim $ 2 \ifmmode\times\else\texttimes\fi $10^{8}$. The on-resonance optical depth ($b_{0}$) of the MOT, defined as $b_{0} = \frac{3 N}{(k \sigma)^{2}}$ where $k$ is the wavenumber of the light resonant to the ${}^{1}S_{0}-{}^{3}P_{1}$ transition, is thus 47. Which confirms that the obtained atomic cloud can be used to perform experiments on collective effects in cold atomic samples \cite{Labeyrie,Araujo}. The temperature is about twice the expected from Doppler cooling theory,  which we attribute to multiple scattering in the atomic cloud or transverse spatial intensity fluctuations of the MOT beams as reported in \cite{Salomon, Chaneliere}. Further improvements of our MOT will exploit schemes allowing to increase the number of captured atoms in the blue MOT by shielding them using a resonant beam on the ${}^{1}S_{0}-{}^{3}P_{1}$ transition \cite{Tarruell} and by simultaneously operating the MOT on both transitions in a core-shell geometry \cite{Mun}.

\subsection{Noise reduction and laser frequency instability}

After this illustration of the convenient locking transition, we now turn to a more detailed description of the reduction of noise and frequency instability of the laser by our MTS frequency stabilization scheme. The effectiveness of a laser frequency stabilization system can be analyzed in the frequency domain. The noise of the error signal can be measured and converted to frequency noise power spectral density by using the error signal as a frequency discriminator. The error signal, before being amplified, has a typical peak-to-valley amplitude of \SI{10}{\milli \volt} and its width (from the peak to the valley) is \SI{5}{\mega \hertz}. Hence, the slope of \SI{0.5}{\mega \hertz \per \milli \volt} is used to convert fluctuations of the error signal (in units of mV) to frequency. We note that collisional (pressure) broadening present at room temperature can be reduced in iodine spectroscopy \cite{Shy} by cooling the spectroscopy cell, leading to a narrower frequency discriminator, but this comes along with a reduced amplitude of the signal requiring thus a longer spectroscopic cell.

\begin{figure}[h!]
\centering
\includegraphics[width=0.5\linewidth]{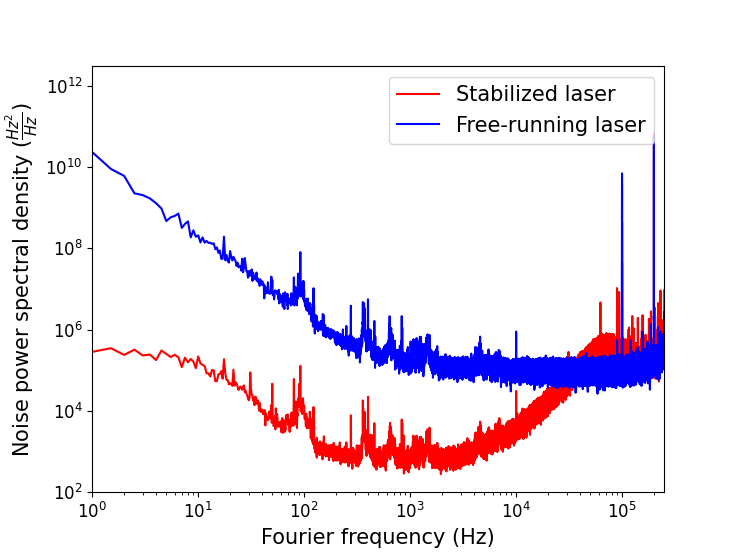}
\caption{Laser noise power spectral density measured from the residual error signal for the free-running (blue) and stabilized (red) laser. The figure shows a noise reduction until \SI{30}{\kilo \hertz}, with a white noise level above \SI{300}{\hertz}  dominated by laser intensity noise. The data shown is an average of 20 data samples.}
\label{NPSD}
\end{figure}

In Fig. \ref{NPSD} we show the noise power spectral density of the laser measured by the error signal noise in blue for the case where the laser is free-running, i.e. no feedback was being sent neither to the piezo nor to the current controllers, and in red for the stabilized laser, when both slow and fast feedback loops are closed. We observe a noise reduction up to \SI{30}{\kilo \hertz}. The white noise part of the free-running spectrum (starting at around \SI{300}{\hertz}) corresponds to the intensity noise level of the laser after the frequency SHG cavity, as reported in \cite{Stuhler}, therefore we attribute this plateau to residual amplitude noise (after PD+ - PD-).
We note that the noise power spectral density shown in  Fig. \ref{NPSD} not only captures the frequency noise of the laser, but residual amplitude noise also contributes to this signal. Nevertheless, the noise reduction shown in Fig. \ref{NPSD} (for frequencies below \SI{1}{\kilo \hertz}) is an indication of the effectiveness of the feedback loop. 

In order to go beyond the analysis of the noise power spectral density, we extract information from the feedback loop by analyzing the Allan deviation of the laser fractional frequency. Although the Allan deviation is still limited by amplitude noise for integration times below  \SI{3}{\milli \second}, the higher values characterize the frequency instability of the laser for averaging times up to  \SI{1}{\second} which is a relevant timescale for the operation of our experiments involving cold atoms. 

\begin{figure}[h!]
\centering
\includegraphics[width=0.5\linewidth]{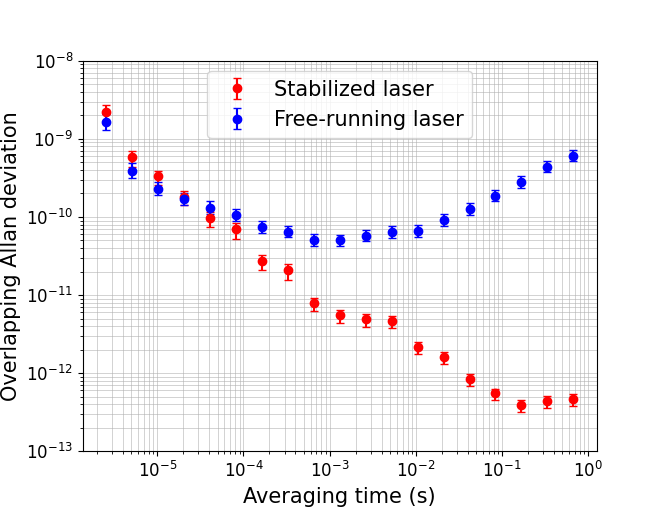}
\caption{Allan deviations of the fractional frequency of the laser calculated by the residual error signal for a free-running (blue) and stabilized (red) laser. The feedback loop starts to be effective at an averaging time of \SI{20}{\micro \second} with an Allan variance of the stabilized laser of 3.9 \ifmmode\times\else\texttimes\fi $10^{-13}$ at \SI{0.17}{\second}.}
\label{Deviation}
\end{figure}

The Allan deviation of an oscillator's fractional frequency is a measure of its frequency instability \cite{Allan1}. We can extract the fractional frequency of the laser from the error signal as the latter is proportional to the frequency fluctuations of the laser, which allows it to be used to characterize the laser's short-term instability \cite{Shang, Zhao, Chen}, as shown in Fig.\ref{Deviation} by the overlapping Allan deviation \cite{Barnes} of the laser's fractional frequency for averaging times ranging from \SI{2.5}{\micro \second} to \SI{0.6}{\second}. The blue points correspond to a free-running laser and the red points to the locked laser with both slow and fast feedback loops closed. For very low averaging times, below \SI{20}{\micro \second}, the laser frequency instability is the same for both cases. However, above \SI{20}{\micro \second} the Allan deviation of the stabilized laser decreases until \SI{0.17}{\second}, while it increases for the free-running case after \SI{1.3}{\milli \second}. We note that the data used to extract the Allan deviation shown in Fig.\ref{Deviation} have been obtained with slightly different parameters of the fast feedback loop than those for the data shown in Fig.  \ref{NPSD}, in particular with an additional slow integrator element of the FALC on the fast circuit branch switched on for the Allan deviation. 
The minimum Allan variance measured is 3.9 \ifmmode\times\else\texttimes\fi $10^{-13}$ at \SI{0.17}{\second}, indicating that the laser frequency stabilization scheme introduced in section 2 can be used for laser cooling of $^{174}$Yb atoms in the ${}^{1}S_{0}-{}^{3}P_{1}$ intercombination line, as shown in the present work.

\section{Conclusion}

In conclusion, we have performed modulation transfer spectroscopy of the P(49)24-1 and R(158)25-0 lines of molecular iodine. The hyperfine components of the R(158)25-0 line are more convenient as a frequency reference for laser cooling of ${}^{174}$Yb, the most naturally abundant ytterbium isotope, because of its smaller frequency separation in comparison to the hyperfine components of the P(49)24-1 line.
We have shown that a laser frequency stabilization technique based on the combination of modulation transfer spectroscopy and balanced detection allows us to stabilize the laser frequency on one of the hyperfine transitions of the R(158)25-0 line, despite its small relative amplitude. The experimental configuration presented can be applied either to enhance the laser frequency stability, as shown in \cite{Mio}, or to achieve laser frequency stabilization in transitions that are relatively weak, as shown in the present work.
We have loaded a MOT, operating in the $^{1}S_{0} - {}^{3}P_{1}$ line of ${}^{174}$Yb, with 2 \ifmmode\times\else\texttimes\fi $10^{8}$ atoms and we have measured a temperature of \SI{60}{\micro \kelvin} by time-of-flight measurements. The on-resonance optical depth of 47 indicates that the obtained atomic cloud can be used to perform experiments on collective effects with narrow atomic transitions, and the results of these experiments will be a benchmark of this system to the study of localization of light by cold atoms in three dimensions \cite{Skipetrov,Kaiser}. The overlapping Allan deviation of the laser fractional frequency was measured for the free-running and the stabilized laser and it emphasizes that the stabilized laser is suitable to perform laser cooling of ${}^{174}$Yb in the narrow intercombination line ($^{1}S_{0} - {}^{3}P_{1}$). Future development of the work presented in this paper will include a deeper study of the impact of laser frequency stabilization onto residual amplitude noise due to the SHG cavity in the laser and possible additional feedforward loops to reduce such amplitude noise. 

This work was performed in the framework of the European project ANDLICA, ERC Advanced grant No. 832219. We also acknowledge support from the project OPTIMAL granted by the European Union by means of the Fond Européen de développement régional, FEDER. We thank Claus Zimmermann and Raphaël Saint-Jalm for fruitful discussions and critical reading of the manuscript.



\begin{thebibliography}{1}

\bibitem{Shimizu}
Takayuki Kurosu and Fujio Shimizu 1990 \textit{Jpn. J. Appl. Phys.} 29 L2127

\bibitem{Ertmer}
Sengstock, K., Sterr, U., Müller, J.H. et al. "Optical Ramsey spectroscopy on laser-trapped and thermal Mg atoms." Appl. Phys. B 59, 99–115 (1994)

\bibitem{Potterveld}
J. R. Guest, N. D. Scielzo, I. Ahmad, K. Bailey, J. P. Greene, R. J. Holt, Z.-T. Lu, T. P. O’Connor, and D. H. Potterveld, "Laser Trapping of $^{225}Ra$ and $^{226}Ra$ with Repumping by Room-Temperature Blackbody Radiation," \textit{Phys. Rev. Lett.} 98, 093001 (2007)

\bibitem{Willmann}
S. De, U. Dammalapati, K. Jungmann, and L. Willmann, "Magneto-optical trapping of barium" \textit{Phys. Rev. A} 79, 041402(R) (2009)

\bibitem{Yabuzaki}
K. Honda, Y. Takahashi, T. Kuwamoto, M. Fujimoto, K. Toyoda, K. Ishikawa, and T. Yabuzaki,"Magneto-optical trapping of Yb atoms and a limit on the branching ratio of the $^{1}P_{1}$ state,"
Phys. Rev. A 59, R934(R) (1999)

\bibitem{Tino}
N. Poli, F.-Y. Wang, M. G. Tarallo, A. Alberti, M. Prevedelli, and G. M. Tino, "Precision Measurement of Gravity with Cold Atoms in an Optical Lattice and Comparison with a Classical Gravimeter," \textit{Phys. Rev. Lett.} 106, 038501 (2011)

\bibitem{Zoller}
Andrew J. Daley, Martin M. Boyd, Jun Ye, and Peter Zoller, "Quantum Computing with Alkaline-Earth-Metal Atoms", \textit{Phys. Rev. Lett.} 101, 170504 (2008)

\bibitem{Gupta} 
Anupriya Jayakumara, Benjamin Plotkin-Swing, Alan O. Jamison, and Subhadeep Gupta, Dual-axis vapor cell for simultaneous laser frequency stabilization on disparate optical transitions,
\textit{Review of Scientific Instruments}
\textbf{86},073115 (2015).

\bibitem{Takahashi}
Uetake, S., Yamaguchi, A., Kato, S. \textit{et al.} High power narrow linewidth laser at 556 nm for magneto-optical trapping of ytterbium. \textit{Appl. Phys. B} \textbf{92}, 33–35 (2008).

\bibitem{Weidemuller}
Luc Couturier, Ingo Nosske, Fachao Hu, \textit{et al}, Rev. Sci. Instrum. \textbf{89}, 043103 (2018)

\bibitem{Cornish} 
A Guttridge \textit{et al} 2016 \textit{J. Phys. B: At. Mol. Opt. Phys.} \textbf{49} 145006.

\bibitem{Fortson}
R. Maruyama, R. H. Wynar, M. V. Romalis, A. Andalkar, M. D. Swallows, C. E. Pearson, and E. N. Fortson
\textit{Phys. Rev. A} \textbf{68}, 011403(R) (2003).

\bibitem{Lu}
Z. Xiong, Y. Long, H. Xiao, X. Zhang, L. He, and B. Lü, \textit{Chin. Opt. Lett.}
\textbf{9}, 041406 (2011).



\bibitem{Dareau}
 Dareau, A. (2015). \textit{Manipulation cohérente d’un condensat de Bose-Einstein d’ytterbium
sur la transition "d’horloge" : de la spectroscopie au magnétisme artificiel.} PhD
Thesis. École Normale Superieure.

 \bibitem{Ducloy}
R. K. Raj, D. Bloch, J. J. Snyder, G. Camy, and M. Ducloy
\textit{Phys. Rev. Lett.} \textbf{44}, 1251 (1980).

\bibitem{DucloyMTS}
G. Camy, Ch. J. Bordé, and M. Ducloy, “Heterodyne saturation spectroscopy through frequency modulation of the
saturating beam,” \textit{Opt. Commun.} \textbf{41}(5), 325–330 (1982).

\bibitem{Himsworth}
Matthew Aldous, Jonathan Woods, Andrei Dragomir, Ritayan Roy, and Matt Himsworth, "Carrier frequency modulation of an acousto-optic modulator for laser stabilization," \textit{Opt. Express} 25, 12830-12838 (2017)
\bibitem{Mio}
Teruhito Hori, Akito Araya, Shigenori Moriwaki, and Norikatsu Mio, "Formulation of frequency stability limited by laser intrinsic noise in feedback systems," Appl. Opt. 48, 429-435 (2009)



\bibitem{Hong}
Y. Tanabe, Y. Sakamoto, T. Kohno, D. Akamatsu, and F. Hong, "Frequency references based on molecular iodine for the study of Yb atoms using the 1S0 – 3P1 intercombination transition at 556 nm," \textit{Opt. Express}  30, 46487-46500 (2022).

\bibitem{DucloyShift}
J. J. Snyder, R. K. Raj, D. Bloch, and M. Ducloy, "High-sensitivity nonlinear spectroscopy using a frequency-offset pump," \textit{Opt. Lett.} 5, 163-165 (1980).

\bibitem{Natarajan}
Kanhaiya Pandey, Alok K. Singh, P. V. Kiran Kumar, M. V. Suryanarayana, and Vasant Natarajan, "Isotope shifts and hyperfine structure in the 555.8-nm ${^{1}S}_{0}\ensuremath{\rightarrow}{^{3}P}_{1}$ line of Yb,"
\textit{Phys. Rev. A} 80, 022518 (2009).

\bibitem{Mcferran}
Peggy E. Atkinson, Jesse S. Schelfhout, and John J. McFerran, "Hyperfine constants and line separations for the $^{1}S_{0}\ensuremath{-}^{3}P_{1}$ intercombination line in neutral ytterbium with sub-Doppler resolution,"
\textit{Phys. Rev. A} 100, 042505 (2019).

\bibitem{Cho}
Heung-Ryoul Noh, Sang Eon Park, Long Zhe Li, Jong-Dae Park, and Chang-Ho Cho, "Modulation transfer spectroscopy for 87Rb atoms: theory and experiment," \textit{Opt. Express} 19, 23444-23452 (2011)

\bibitem{Turner}
V. Negnevitsky and L. D. Turner, "Wideband laser locking to an atomic reference with modulation transfer spectroscopy," \textit{Opt. Express} 21, 3103-3113 (2013)

\bibitem{Lin}
Zhang Zhang, Xiaolong Wang, and Qiang Lin, "A novel way for wavelength locking with acousto-optic frequency modulation," \textit{Opt. Express} 17, 10372-10377 (2009)

\bibitem{Skipetrov}
S.E. Skipetrov and I.M. Sokolov, "Magnetic-Field-Driven Localization of Light in a Cold-Atom Gas,"
Phys. Rev. Lett. 114, 053902 (2015)

\bibitem{Letellier}
Hector Letellier, Álvaro Mitchell Galvão de Melo, Anaïs Dorne and Robin Kaiser, "Loading of a large Yb MOT on the 1S0 →1 P1 transition",	\textit{arXiv:2308.00387} (2023)

\bibitem{Shaffer}
K. R. Overstreet, P. Zabawa, J. Tallant, A. Schwettmann, and J. P. Shaffer, "Multiple scattering and the density distribution of a Cs MOT," \textit{Opt. Express} 13, 9672-9682 (2005)

\bibitem{Gawlik}
Tomasz M Brzozowski et al 2002 \textit{J. Opt. B: Quantum Semiclass. Opt.} 4 62

\bibitem{Labeyrie}
G. Labeyrie, F. de Tomasi, J.-C. Bernard, C. A. Müller, C. Miniatura, and R. Kaiser, "Coherent Backscattering of Light by Cold Atoms,"
\textit{Phys. Rev. Lett.} 83, 5266 (1999)

\bibitem{Araujo}
Michelle O. Araújo, Ivor Krešić, Robin Kaiser, and William Guerin, "Superradiance in a Large and Dilute Cloud of Cold Atoms in the Linear-Optics Regime," \textit{Phys. Rev. Lett.} 117, 073002 (2016)

\bibitem{Salomon}
Drewsen, M., Laurent, P., Nadir, A. et al. Investigation of sub-Doppler cooling effects in a cesium magneto-optical trap. \textit{Appl. Phys. B} 59, 283–298 (1994).

\bibitem{Chaneliere}
Thierry Chanelière, Jean-Louis Meunier, Robin Kaiser, Christian Miniatura, and David Wilkowski, "Extra-heating mechanism in Doppler cooling experiments," \textit{J. Opt. Soc. Am. B} 22, 1819-1828 (2005)

\bibitem{Tarruell}
Jonatan Höschele, Sandra Buob, Antonio Rubio-Abadal, Vasiliy Makhalov, and Leticia Tarruell, "Atom-Number Enhancement by Shielding Atoms From Losses in Strontium Magneto-Optical Traps", \textit{Phys. Rev. Applied} 19, 064011 (2023)

\bibitem{Mun}
Jeongwon Lee, Jae Hoon Lee, Jiho Noh, and Jongchul Mun, "Core-shell magneto-optical trap for alkaline-earth-metal-like atoms", \textit{Phys. Rev. A} 91, 053405 (2015)

\bibitem{Shy}
Hui-Mei Fang, S.C. Wang, Jow-Tsong Shy, “Pressure and power broadening of the a10 component of R(56) 32-0 transition of molecular iodine at 532nm,” \textit{Opt. Commun.} \textbf{257}, 76–83 (2006)

\bibitem{Stuhler}
Ulrich Eismann, Martin Enderlein, Konstantinos Simeonidis, Felix Keller, Felix Rohde, Dmitrijs Opalevs, Matthias Scholz, Wilhelm Kaenders, Jürgen Stuhler, "Active and passive stabilization of a high-power UV frequency-doubled diode laser",	\textit{Conference on Lasers and Electro-Optics, OSA Technical Digest (online)} (Optica Publishing Group, 2016), paper JTu5A.65. (2016)

\bibitem{Allan1}
D. W. Allan, "Statistics of atomic frequency standards," \textit{Proceedings of the IEEE}, vol. 54, no. 2, pp. 221-230 (1966)

\bibitem{Shang}
Chang, P., Zhang, S., Shang, H. et al. "Stabilizing diode laser to 1 Hz-level Allan deviation with atomic spectroscopy for Rb four-level active optical frequency standard." \textit{Appl. Phys. B} 125, 196 (2019)

\bibitem{Zhao}
Y T Zhao \textit{et al} 2004 \textit{J. Phys. D: Appl. Phys.} 37 1316

\bibitem{Chen}
Jianxiang Miao, Tiantian Shi, Jia Zhang, and Jingbiao Chen, "Compact 459-nm Cs Cell Optical Frequency Standard with $2.1\ifmmode\times\else\texttimes\fi{{10}^{\ensuremath{-}13}/\sqrt{\ensuremath{\tau}}}$ Short-Term Stability", \textit{Phys. Rev. Applied} 18, 024034 (2022)

\bibitem{Barnes}
D. A. Howe, D. U. Allan and J. A. Barnes, "Properties of Signal Sources and Measurement Methods," \textit{Thirty Fifth Annual Frequency Control Symposium}, Philadelphia, PA, USA, 1981


\bibitem{Kaiser}
Giuseppe Luca Celardo, Mattia Angeli, Robin Kaiser, "Localization of light in subradiant Dicke states: a mobility edge in the imaginary axis",	\textit{arXiv:1702.04506} (2017)




\end{thebibliography}


\end{document}